\documentclass[conference]{IEEEtran}
\usepackage{graphicx,times,cite,amsmath, amssymb, epsfig, array, mathtools, epstopdf,makecell, algorithm, algorithmic,dsfont,color,xfrac}
\newlength{\figwidth}
\begin{document}
\newcommand{\new}[1]{\textcolor{blue}{#1}}	
\setlength{\pdfpagewidth}{8.5in}
\setlength{\pdfpageheight}{11in}

\newtheorem{theorem}{Theorem}
\newtheorem{corollary}{Corollary}
\newtheorem{lemma}{Lemma}
\newtheorem{proposition}{Proposition}
\newtheorem{condition}{Condition Set}
\newtheorem{prob}{Problem}

\IEEEoverridecommandlockouts

\title{Resource Allocation in Dynamic DF Relay for SWIPT Network with Circuit Power Consumption}


\author{Bhathiya Pilanawithana, Saman Atapattu and Jamie Evans\\
\IEEEauthorblockA{Department of Electrical and Electronic Engineering, University of Melbourne, Parkville, VIC 3010, Australia.\\
Email: mpilanawitha@student.unimelb.edu.au;\{saman.atapattu, jse\}@unimelb.edu.au}
}

\maketitle

\begin{abstract}
This paper considers simultaneous wireless information and power transfer (SWIPT) over a dual-hop dynamic decode-and-forward (DF) relay network with the power-splitting (PS) energy harvesting protocol at the relay. The circuit power consumption (CPC), which includes power requirements for both decoding and encoding circuits, is considered at the relay. For a rate-dependent linear CPC model, we formulate an optimization problem to decide the optimal throughput, PS ratio, relay transmit power and  time ratio for the source to relay transmission. Although the resultant optimization problem is non-convex, we derive an efficient optimization algorithm, requiring significantly less floating point operations than an interior point method. Finally, we present numerical results which lead to some interesting insights for system design.
\end{abstract}

\section{Introduction}
Communication nodes in wireless sensor networks (WSNs) and Internet of Things (IoT) are typically powered by individual power supplies \cite{Guo2018guotcom}. Due to inconsistent availability and/or implementation overhead for fixed and ambient energy sources such as solar, wind or vibration, wireless energy transfer (WET) and ambient backscatter communications have been introduced as promising techniques which require low-cost modifications to existing communication circuitry \cite{Zhou,AmanthiICC2016,Zhao2018coml,Zhao19tvt}.
Since the same radio frequency (RF) signal can carry both energy and information, 
WET approach is known as simultaneous wireless information and power transfer (SWIPT). Cooperative communication with relay nodes is a vital ingredient in WSNs to improve connectivity and energy efficiency \cite{Erkip}. Further, relaying is an effective technique to improve the wireless connectivity by helping to extend the wireless network coverage without a need to deploy wired backhaul facilities, e.g.,  \cite{Atapattu2019tcom} and references therein. Due to this reason, SWIPT in relay networks has gained much attention recently  as detailed below.

For SWIPT, the time-switching (TS), power-splitting (PS) and hybrid WET protocols are introduced for both amplify-and-forward (AF) and decode-and-forward (DF) relay networks in \cite{Nasir1,ZDing,Atapattu15icc, Saman3,Fan2018acc}, where the optimal PS and TS ratios are derived to achieve the maximum performance with negligible circuit power consumption (CPC). 
However, when the source node is the only node with a power supply, the benefits of using a relay may rapidly decay with CPC 
\cite{Sun2012}. The CPC can be modeled as an information rate dependent \cite{Isheden,Bai,GFeng,ZZhong},  transmit-power (i.e., power used for the information transmission) dependent \cite{Veciana}, or as a fixed power \cite{Alves}. 

Despite a lot of literature on WET, analysis considering CPC appears to be  lacking. In \cite{ZZhong}, optimal energy beamforming and time assignment is derived considering a rate-dependent CPC model in a wireless powered sensor network in which each sensor node directly transmits information to a destination node. However, this work is not for relay networks. In \cite{Bai}, energy and information beamforming is jointly optimized considering constant CPC in each wireless powered sensor that directly transmits information to a destination node. In \cite{JXu}, a transmit power-dependent CPC model is used to derive optimal power allocation for a transmitting node powered by energy harvesting from ambient energy sources which transmits information directly to a destination node. However, to the best of our knowledge, no work has considered the  SWIPT relay networks.

In addition, most of the work assumes static and equal time durations for source to relay and relay to destination transmissions in the DF relay.  However, \cite{VTarokh,KVaranasi} shows that the DF relay with dynamic time durations which are dependent on the channels, outperforms the static case. The diversity order and the sum information rate of dynamic DF relays are calculated in \cite{RSchober}. The dynamic DF concept is applied to a two-way SWIPT relay in \cite{TPDo} to derive the optimal outage probability for PS and TS protocols.   

To the best of our knowledge, this is the first work which considers an optimization framework for resource allocation considering rate-dependent CPC at a dynamic DF relay. In particular, this paper analyzes a dual-hop dynamic DF relay network with the PS protocol by considering CPC at the relay which includes power requirements for both decoding and encoding circuits. We use the information rate dependent linear CPC model as our target is to maximize the information rate. Then, we formulate the optimal resource allocation problem to jointly optimize throughput, PS ratio, power allocation at the relay and the time ratio for source to relay transmission.  We show that the problem is non-convex and propose an efficient sub-optimal algorithm which achieve near-optimal performance. 

The rest of this paper is organized as follows. Section~\ref{system_model} presents the system model. Section~\ref{opt_prob} solves optimization problem for the resource allocation and provides low-complex near-optimal resource allocation algorithm. Section~\ref{num_res} presents numerical and simulation results. The concluding remarks are in Section~\ref{con}  followed by the respective proof in Appendix.


\section{System Model}\label{system_model}
\begin{figure}
	\centering
	\includegraphics[width=1\figwidth]{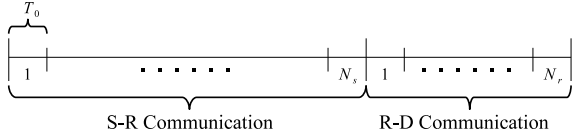}
	\caption{Transmission frame structure}
	\label{Frame}
\end{figure}
\subsection{Network Model}
We consider a half-duplex wireless relay network where a source node ($S$) communicates with a destination node ($D$), via a DF relay node ($R$). All three nodes operate with same fixed sampling time $T_0$. We assume that the direct link between $S$ and $D$ is not available due to a blockage. As shown in the Fig.~\ref{Frame},  $S$ transmits to $R$ using first $N_s$ samples with transmit power $Q$ and $R$ transmits to $D$ using the remaining $N_r$ samples with transmit power $P_t$. The ratio of $S-R$ transmit time compared to the total block time is denoted by $\theta=\sfrac{N_s}{\left(N_s+N_r\right)}$. The channel coefficients of the $S$ to $R$ ($S-R$) and $R$ to $D$ ($R-D$) channels are $h_1$ and $h_2$, respectively, which stay constant during the total block time $\left(N_s+N_r\right)T_0$. The source generates information with rate $\tau$\, (bits/s). The relay uses $\lambda$ and $\left(1-\lambda\right)$ portions of the received signal for information and energy harvesting respectively using the PS protocol. At the beginning of each slot, channel state information (CSI), i.e, $h_1$ and $h_2$, is available at a decision node which chooses $\tau$, $\lambda$, $P_t$ and $\theta$, and passes them to all three nodes (anyone of the three nodes may be considered as the decision node).

\subsection{Analytical Model}
\begin{figure}
	\centering
	\includegraphics[width=1.1\figwidth]{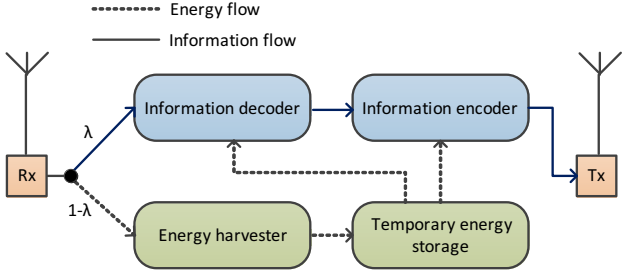}
	\caption{Block diagram of the relay}
	\label{Relay_Block}
\end{figure}
Signal-to-noise ratios of $S-R$ and $R-D$ links can be given respectively as $\frac{Q g_1 \lambda}{\left(1+\lambda\right) \sigma^2}$ and $\frac{P_t  g_2}{2 \sigma^2}$, \cite{Saman3}. Therefore, the maximum amount of information, that can be transferred in $S-R$ and $R-D$ links can be written as
\begin{equation}\label{constrain_tau_C_sr}
B_{sr}  = N_s \, \text{log}_2 \left(1+\frac{Q g_1 \lambda}{\left(1+\lambda\right) \sigma^2}\right) \,\text{bits}\ ,
\end{equation}
\begin{equation}\label{constrain_tau_C_rd}
B_{rd}  = N_r \, \text{log}_2 \left(1+\frac{P_t  g_2}{2 \sigma^2}\right) \,\text{bits}\ ,
\end{equation}
where $g_1=|h_1|^2$, $g_2=|h_2|^2 $ and $\sigma^2$ is the noise \break power. Since the source generates $\left(N_s+N_r\right)T_0 \tau$ information  bits at the beginning of each block, we have $\left(N_s+N_r\right)T_0 \tau \leqslant \text{min}\left(B_{sr},B_{rd}\right)$. This can be given as
\begin{align}\label{tau_const}
\tau\leqslant\frac{ B_{sr}}{\left(N_s+N_r\right)T_0} \,\,\text{and} \, \tau\leqslant \frac{ B_{rd}}{\left(N_s+N_r\right)T_0} \ .
\end{align}

As shown in Fig.~\ref{Relay_Block}, the relay consists of information decoding, energy harvesting, information encoding circuits, and also a temporary energy storage (e.g. capacitor). Since $\left(N_s+N_r\right)T_0 \tau$ information bits sent to the decoding circuits during a $T_0 N_s$ period, the effective information rate at the relay decoding circuits, is $\frac{\tau}{\theta}$. Similarly, effective information rate at the relay encoding circuits, is $\frac{\tau}{\left(1-\theta\right)}$. As CPC is modeled as an information
rate dependent linear model [8], CPCs at decoding, $P_{dec}$, and
encoding, $P_{enc}$, can be given, respectively, as
\begin{align}
&P_{dec}=P_d+ \epsilon_d \frac{\tau}{\theta} \,\,\text{and} \,
P_{enc}=P_t+P_e+ \epsilon_e \frac{\tau}{\left(1-\theta\right)} \ , \label{P_dec}
\end{align}
where $ \epsilon_d,\,\epsilon_e>0$ are unit rate dynamic energy consumption at decoder and encoder, respectively; and $P_d$ and $P_e$ are static power consumption at the decoder and the encoder, respectively. 
The total energy used by the relay can not be larger than the total harvested energy. Thus, with the aid of (\ref{P_dec}),  we can write
\begin{multline}\label{constrain_total_harvested}
\left(P_d+ \epsilon_d \sfrac{\tau}{\theta}\right)T_0 N_s + \left(P_t+P_e+ \epsilon_e \sfrac{\tau}{\left(1-\theta\right)}\right) T_0 N_r \\ \leqslant  \eta Q g_1 \left(1-\lambda\right) T_0 N_s \ ,
\end{multline}
where $E_h =  \eta Q g_1 \left(1-\lambda\right) T_0 N_s$ is the  total harvested energy in $R$. 
This shows that optimal choice of source information rate, PS ratio, relay transmit power and $S-R$ transmit time ratio depends on the CPC.

\section{Resource Allocation}\label{opt_prob}
Our objective is to design $\lambda$, $P_t$ and $\theta$, in order to maximize the source information rate, $\tau$, for given values of $g_1$ and $g_2$. The optimization problem can be given as
\begin{subequations}\label{PS_opti_prob}
	\begin{align}
	&\underset{\tau,\  \lambda,\ P_t,\ \theta}{\text{max}}  \quad  \tau\\
    &\quad \ \ \text{s.t.} \quad \ \tau - \frac{\theta}{T_0} \ \text{log}_2 \left(1+\frac{Q g_1 \lambda}{\left(1+\lambda\right) \sigma^2}\right) \leqslant 0 \label{constraint1}\\
  	&\quad\quad\quad\quad  \tau - \frac{\left(1-\theta\right)}{T_0}\text{log}_2 \left(1+\frac{P_t  g_2}{2 \sigma^2}\right) \leqslant 0 \label{constraint2}\\
	&\quad P_d \theta +P_e\left(1-\theta\right)+\left(\epsilon_d+\epsilon_e\right)\tau+P_t\left(1-\theta\right) \nonumber\\ &\quad\qquad \qquad \qquad \qquad \quad\quad\quad \ \leqslant  \eta Q g_1 \left(1-\lambda\right) \theta \label{constraint4}\\
	&\quad\quad\quad\quad 0\leqslant \lambda, \ 0\leqslant P_t, \ 0 < \tau , \ 0<\theta<1 \ . \label{constraint5}
	\end{align}
\end{subequations}
The constraint \eqref{constraint1} and \eqref{constraint2} comes from \eqref{tau_const}, while constraint \eqref{constraint4} comes from dividing \eqref{constrain_total_harvested} by $\left(N_s+N_r\right)T_0$.
The condition $\lambda\leqslant 1$ is explicitly satisfied in (\ref{constraint4}) as $\tau\geqslant 0$ and $P_t\geqslant 0$. The optimal solution is denoted by $\left(\tau^*,P^*_t,\lambda^*,\theta^*\right)$.

\subsection{Feasibility of the Resource Allocation}
If the maximum achievable  of the right hand side of the inequality \eqref{constraint4} is smaller than the minimum achievable of the left hand side, the feasible set of the optimization problem \eqref{PS_opti_prob} is empty, in which case the resource allocation problem is infeasible. Therefore, for a fixed $\theta$, a condition for feasibility can be written as $P_d \theta +P_e\left(1-\theta\right) <  \eta Q g_1 \theta$. The set of all $\theta$ that satisfy this condition, can be given by $\left(\theta_0,1\right)$, where \break $\theta_0=\sfrac{P_e}{\left(\eta Q g_1 + P_e - P_d\right)}$. Since $\theta \in (0,1)$, the feasible set $\left(\theta_0,1\right)$ is empty if $\eta Q g_1 \leqslant P_d $. In addition, $g_2>0$ is required for $\tau>0$ in \eqref{constraint2}.  Thus, we solve the optimization problem only when $\eta Q g_1 > P_d $ and $g_2>0$. The optimization problem is non-convex, but we next discuss an efficient method to find a near optimal solution.

\subsection{Near Optimal Method for the Resource Allocation}
By noticing that \eqref{PS_opti_prob} is convex for fixed $\theta$, we first solve the optimization problem for fixed $\theta=\bar{\theta} \in (\theta_0,1)$ and use a one-dimensional grid search to get the near optimal overall solution. 
\begin{proposition}\label{prop2}
	When $P_d \bar{\theta} +P_e\left(1-\bar{\theta}\right) <  \eta Q g_1 \bar{\theta}$ and $g_2>0$, optimal decision variables for fixed $\theta=\bar{\theta}$ denoted by $\left(\bar{\tau}, \bar{\lambda}, \bar{P_t} \right)$, satisfy
	\begin{equation*}
	\bar{\lambda}= f_1\left(\bar{P_t},\bar{\theta}\right) = \frac{\sigma^2}{ \left(\frac{Q g_1}{\left(1+\frac{\bar{P_t} g_2}{2 \sigma^2}\right)^{\frac{1-\bar{\theta}}{\bar{\theta}}}-1} - \sigma^2\right)} \ ,  
	\end{equation*}
	\begin{equation*}
	\bar{\tau}=f_2\left(\bar{P_t},\bar{\theta}\right)=\frac{\left(1-\bar{\theta}\right)}{T_0}\text{log}_2\left(1+\frac{\bar{P_t} g_2}{2 \sigma^2}\right) \ . 
	\end{equation*}
	Furthermore, $\bar{P_t}\in \left[0,c\right]$ is uniquely determined by
	\begin{equation*}
	f_{\bar{\theta}}\left(\bar{P_t}\right)=0 \ ,
	\end{equation*}
	where 
	\begin{multline*}
	f_{\bar{\theta}}\left(P_t\right)=\eta Q g_1 \theta \left(1-f_1\left(P_t,\bar{\theta}\right)\right)-\bar{\theta} P_d-\left(1-\bar{\theta}\right)P_e - \\ \left(\epsilon_d+\epsilon_e\right)f_2\left(P_t,\bar{\theta}\right)-\left(1-\bar{\theta}\right)P_t
	\end{multline*}
	and
	\begin{equation*}
	c=\frac{2 \sigma^2}{g_2}\left(\left(1+\frac{Q g_1}{2 \sigma^2}\right)^{\frac{\bar{\theta}}{\left(1-\bar{\theta}\right)}}-1\right) \ .
	\end{equation*}
\end{proposition}
\begin{IEEEproof}
	See Appendix~A
\end{IEEEproof}
It can be easily shown that $f_{\bar{\theta}}\left(P_t\right)$ is strictly decreasing in $[0,c]$. Together with $f_{\bar{\theta}}\left(0\right)>0$ and $f_{\bar{\theta}}\left(c\right)\leqslant 0$,  this ensures that $\bar{P_t}$ is unique. Thus, any root finding algorithm can be used to solve $f_{\bar{\theta}}\left(\bar{P_t}\right)=0$. To get a near-optimal solution to \eqref{PS_opti_prob}, we perform a simple grid search along $\theta\in (\theta_0,1)$ with $n$ number of grid levels using the following algorithm.
\begin{algorithm}
	\caption{Near Optimal Resource Allocation}
	\label{resource}
	\begin{algorithmic}
		\STATE 
		\STATE \textbf{Input:} grid levels $n>0$, $g_1$,  $g_2$,  $Q$,  $\sigma^2$, $T_0$
		\STATE $\left(\tau^*,P^*_t,\lambda^*,\theta^*\right) \leftarrow \left(0,0,0,0\right)$
		\IF{$\eta Q g_1 >P_d$\ and $g_2>0$}
		\FOR{$i=1,\cdots,n$}
		\STATE $\bar{\theta} \leftarrow \theta_0+\sfrac{i\left(1-\theta_0\right)}{\left(n+1\right)}$
		\STATE Solve $f_{\bar{\theta}}\left(\bar{P_t}\right)=0$ where $\bar{P_t} \in \left[0,c\right]$
		\STATE $\bar{\tau}=\frac{\left(1-\bar{\theta}\right)}{T_0}\text{log}_2\left(1+\frac{\bar{P_t} g_2}{2 \sigma^2}\right)$
		\STATE $\bar{\lambda}= \frac{\sigma^2}{ \left(\frac{Q g_1}{\left(1+\frac{\bar{P_t} g_2}{2 \sigma^2}\right)^{\frac{1-\bar{\theta}}{\bar{\theta}}}-1} - \sigma^2\right)}$
		\IF{$\bar{\tau}>\tau^*$}
		\STATE $\left(\tau^*,P^*_t,\lambda^*,\theta^*\right) \leftarrow \left(\bar{\tau},\bar{P_t},\bar{\lambda},\bar{\theta}\right)$
		\ENDIF
		\ENDFOR
		\ENDIF
		\RETURN $\left(\tau^*,P^*_t,\lambda^*,\theta^*\right)$
	\end{algorithmic}
\end{algorithm}

\subsection{A Lower Bound for Maximum Source Information Rate}
Feasible set of the problem is defined by the constraints of the problem \eqref{PS_opti_prob}. A lower bound can be obtained by considering a suitable element of the feasible set. To this end, we first set $\theta=\sfrac{1}{2}$ and by using the Proposition~\ref{prop2} we can write
\begin{equation}\label{lo_f1}
\lambda= f_1\left(P_t,\frac{1}{2}\right) = \frac{P_t g_2}{2 Q g_1 - P_t g_2} \ ,  
\end{equation}
\begin{equation}\label{lo_f2}
\tau=f_2\left(P_t,\frac{1}{2}\right)=\frac{1}{2 T_0}\text{log}_2\left(1+\frac{P_t g_2}{2 \sigma^2}\right) \ , 
\end{equation}
where $P_t$ is in $[0,\sfrac{Q g_1}{g_2}]$. The solution to $f_{\theta}\left(P_t\right)=0$ can not be obtained in closed form due to the rational and the log function in \eqref{lo_f1} and \eqref{lo_f2}, respectively. Thus, we use a suitable $P_t\in [0,\sfrac{Q g_1}{g_2}]$ that satisfy  constraint \eqref{constraint4}, which can be used to obtain a lower bound for $\tau^*$. Thereupon, we notice that $f_2\left(P_t,\sfrac{1}{2}\right)$ is a concave function and we use the tangent at  $f_2\left(0,\sfrac{1}{2}\right)$ to obtain an upper bound for the function $f_2\left(P_t,\sfrac{1}{2}\right)$. This can be written as,
\begin{equation}\label{f_2_up}
f_2\left(P_t,\frac{1}{2}\right) \leqslant \frac{g_2 P_t}{4 T_0 \sigma^2}
\end{equation}
Moreover, we also notice that the function $\left(1-f_1\left(P_t,\frac{1}{2}\right)\right)$ is concave. Thus, using the straight line between $f_1\left(0,\sfrac{1}{2}\right)$ and $f_1\left(\sfrac{Q g_1}{g_2},\sfrac{1}{2}\right)$ we can write
\begin{equation}\label{f_1_low}
\eta Q g_1  \left(1-f_1\left(P_t,\frac{1}{2}\right)\right) \geqslant \eta  \left(Q g_1 - P_t g_2\right) 
\end{equation}
In order to find a lower bound for $\tau^*$, functions $f_2\left(P_t\right)$ and $f_1\left(P_t\right)$ are upper and lower bounded linear functions of $P_t$ to assume that CPC requires more power than the model in \eqref{P_dec} and the total harvested energy is smaller than $\eta Q g_1 \left(1-\lambda\right) T_0 N_s$. Therefore, if
\begin{multline}\label{low_pt}
 \frac{P_d}{2}  +\frac{P_e}{2}+\left(\epsilon_d+\epsilon_e\right)\frac{g_2 P_t}{4 T_0 \sigma^2}+\frac{P_t}{2} = \frac{\eta \left(Q g_1 - P_t g_2\right)}{2} \, ,
\end{multline}
then with the aid of \eqref{f_2_up} and \eqref{f_1_low} we have the constraint \eqref{constraint4} satisfied. In comparison to $f_{\theta}\left(P_t\right)=0$ the equation \eqref{low_pt} is a solution to a linear function of $P_t$, which can be simplified to give
\begin{equation*}
P_t = \frac{\eta Q g_1 - P_d-P_e}{1+\eta g_2+ \frac{\left(\epsilon_d+\epsilon_e\right) g_2}{2 T_0 \sigma^2}}\, ,
\end{equation*}
which can take a negative value if $\eta Q g_1 < P_d+P_e$. Therefore, by substituting in \eqref{lo_f2}, a lower bound for $\tau^*$ can be written as,
\begin{equation}\label{lower}
\tau^* \geqslant \max\left[0,\frac{1}{2 T_0}\text{log}_2\left(1+\frac{\eta Q g_1 - P_d-P_e}{\frac{2 \sigma^2}{g_2}+2\sigma^2 \eta+ \frac{\left(\epsilon_d+\epsilon_e\right)}{T_0}}\right)\right]
\end{equation}

\section{Numerical and Simulation Results}\label{num_res}
In this section, we use $Q=500$\,mW, $\sigma^2 = 10$\,mW, $T_0=500$\,$\mu$s, $\eta=0.8$ and $n=500$.

\begin{figure}[t!]
	\centering
	\includegraphics[width=1\figwidth]{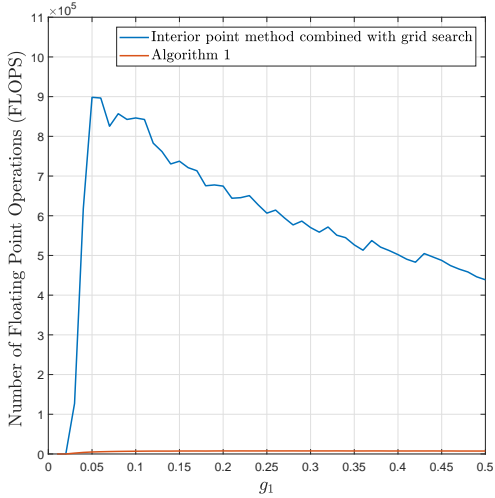}
	\caption{Variation of the total number of FLOPS with $g_1$ when $P_e,P_d=10$\,mW, $\epsilon_e+\epsilon_d=0.1$\,mW/bits/s, $g_2=0.3$}
	\label{FLOPS}
\end{figure}
\begin{figure}[t!]
	\centering
	\includegraphics[width=0.95\figwidth]{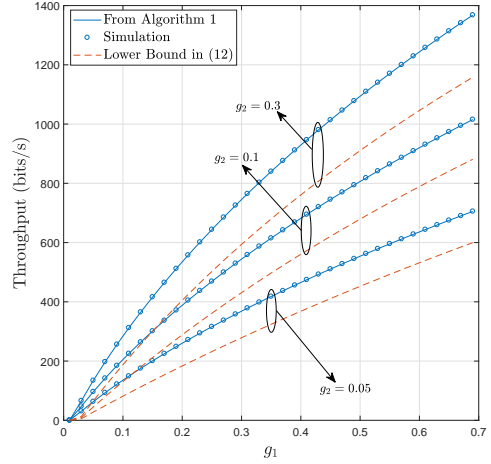}
	\caption{Variation of the throughput with $g_1$, when $P_e,P_d=10$\,mW, $\epsilon_e+\epsilon_d=0.1$\,mW/bits/s and $g_2=0.05$, $0.1$ and $0.3$.}
	\label{Simulation_vs_analytical}
\end{figure}

If the optimization problem (\ref{PS_opti_prob}) for fixed $\theta$ is solved using an interior-point method with Newton step, it involves a matrix inversion of a $3\times 3$ matrix in every iteration \cite{Boyed}. If Gaussian elimination is used for the matrix inversion, total number of floating point operations (FLOPS) per iteration is $27$  for that matrix inversion. In contrast, the bisection method requires only $3$ FLOPS per iterations to solve $f_{\bar{\theta}}\left(\bar{P_t}\right)=0$ in proposition~\ref{prop2}. Fig~\ref{FLOPS} shows the variation of the total number of FLOPS (=~number of iterations\,$\times$\,FLOPS per iteration,$\times$\,$n$) with $g_1$ when $P_e,P_d=5$\,mW, $\epsilon_e+\epsilon_d=10$\,mW/bits/s/Hz, $g_2=0.1$. As shown in the figure, the total FLOPS is at least $60$ times higher in the interior-point method. Due to this calculation efficiency, we use Algorithm~1 to generate the rest of the numerical results. 

\begin{figure}[t!]
	\centering
	\includegraphics[width=0.95\figwidth]{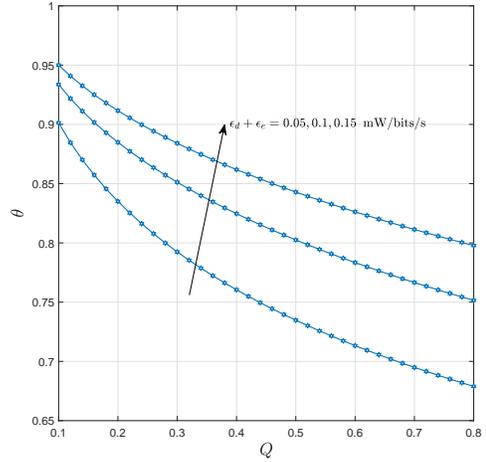}
	\caption{Variation of the optimal $\theta$ with $Q$, when $\epsilon_d+\epsilon_e=0.05,0.1,0.15$\,mW/bits/s, $P_d,P_e=10$\,mW $g_1,g_2=0.4$.}
	\label{op_alpha}
\end{figure}
\begin{figure}[t!]
	\centering
	\includegraphics[width=0.95\figwidth]{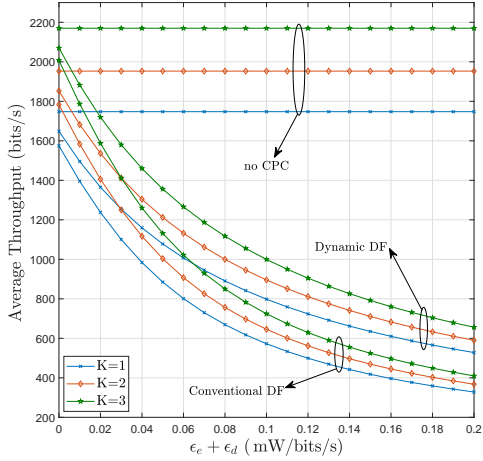}
	\caption{Variation of the average throughput with $\epsilon_e+\epsilon_d$, when $P_d, P_e=10$\,mW, $Q=500$\,mW and Rice factor $K=1,2$ and $3$.}
	\label{Dynamic_pwr}
\end{figure}
Fig.~\ref{Simulation_vs_analytical} shows the variation of the optimal throughput with $g_1$ when $P_e,P_d=5$\,mW, $\epsilon_e+\epsilon_d=0.1$\,mW/bits/s for $g_2=0.05$, $0.1$ and $0.3$. The lower bound in \eqref{lower} for maximum source information rate is also shown in the figure. The  simulation results for each pair of  $g_1$ and $g_2$ is obtained by solving the optimization Problem~(\ref{PS_opti_prob}) using an interior point method for fixed $\theta$ combined with a grid search in $\theta$. The numerical results obtained by Algorithm~\ref{resource} exactly match with the simulation results, which validates our analysis. As shown in Fig.~\ref{Simulation_vs_analytical}, the lower bound gives a close approximation to maximum source information rate when $g_1$ and $g_2$ are small.

Fig.~\ref{op_alpha} shows the variation of the optimal $\theta$ with $Q$. For smaller $Q$, the optimal $\theta$ is close to $1$. However, when $Q$ is large, the optimal $\theta$ is close to $\sfrac{1}{2}$ which make the conventional DF relay performance close to dynamic DF relay performance. This due to the fact that, large received power enables the relay to harvest sufficient amount of energy with $\theta=\sfrac{1}{2}$.

\begin{figure}[t!]
	\centering
	\includegraphics[width=0.95\figwidth]{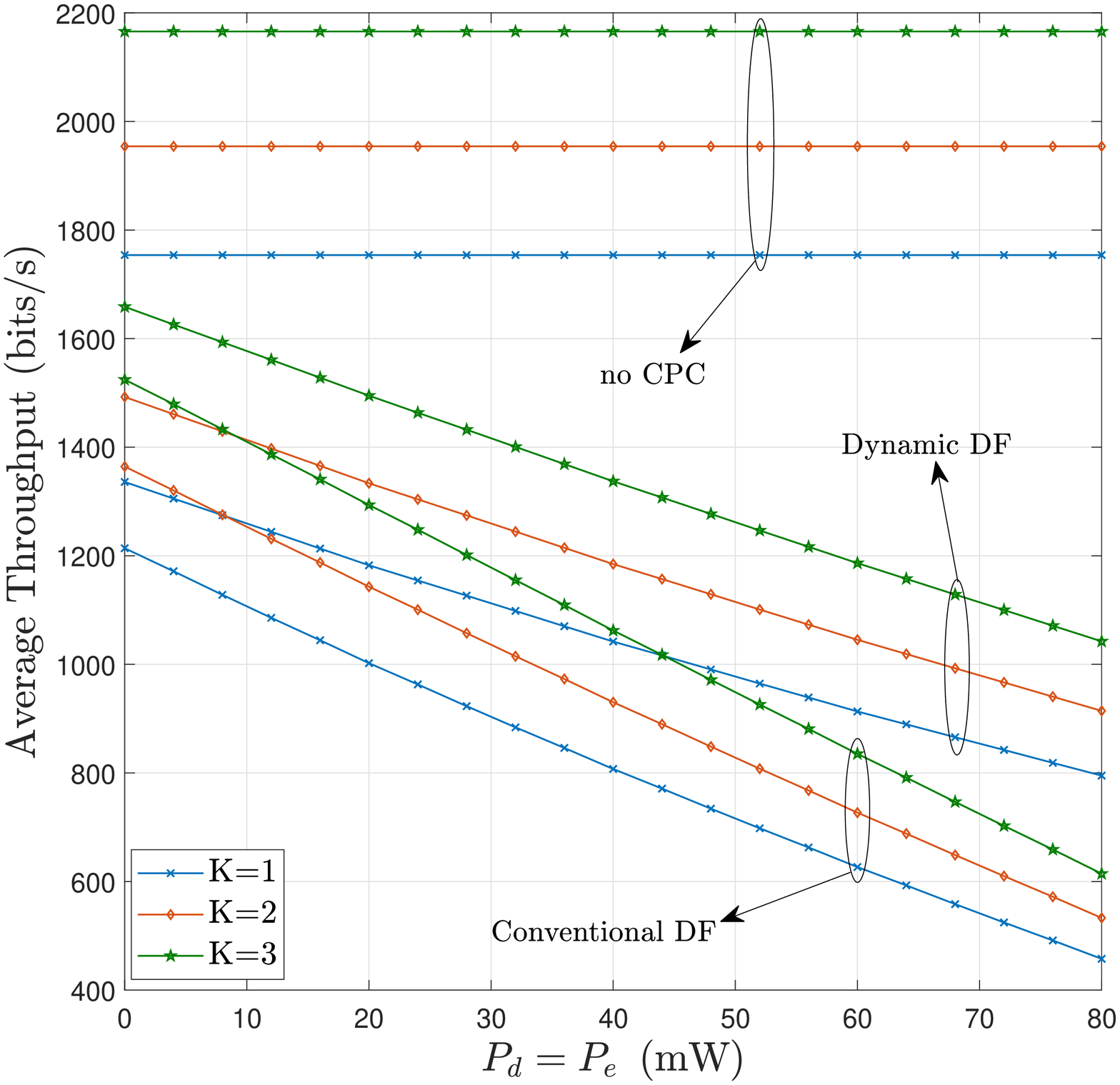}
	\caption{Variation of the average throughput with $P_e=P_d$, when $\epsilon_e+\epsilon_d=0.03$\,mW/bits/s, $Q=500$\,mW and Rice factor $K=1,2$ and $3$.}
	\label{Static_pwr}
\end{figure}

Fig.~\ref{Dynamic_pwr} shows the variation of the optimal average throughput with  $\epsilon_e+\epsilon_d$, when $P_d, P_e=10$\,mW. We assume Rician fading channels with Rice factor $K=1,2$ or $3$. For the comparison, we also plot the no CPC case, and the conventional DF relay with $\theta=\sfrac{1}{2}$. The optimal average throughput gradually decreases when $\epsilon_e+\epsilon_d$ increases. This is because the power available for information transmission reduces due to CPC. Dynamic DF relay shows significant performance gains over conventional DF relay when the CPC from $\epsilon_e+\epsilon_d$ is large. Extra design variable $\theta$ in dynamic DF relay provides a trade off between throughput performance and the computational complexity of the resource allocation. Due to a stronger line-of-sight component, the throughput increases when the Rice factor increases. 
Fig.~\ref{Static_pwr} shows the optimal average throughput vs  $P_e=P_d$, when $\epsilon_e+\epsilon_d=0.03$\,mW/bits/s. The optimal throughput gradually decreases when $P_e$ or $P_d$ increases.

\vspace{-1mm}
\section{Conclusion}\label{con}
This paper considers a SWIPT dynamic DF relay network, in which the relay uses the PS energy harvesting protocol. The circuit power consumption at the relay is modeled by an information rate dependent linear model. The throughput is targeted to be maximized by jointly optimizing the PS ratio and relay transmit power and the time ratio for source to relay transmission. This problem is non-convex and an efficient sub-optimal algorithm is given, which achieves a near optimal performance. A closed form expression of a lower bound for maximum source information rate is obtained, which provide deeper insight to the problem structure. Numerical results show that the extra degree of freedom to choose an appropriate transmit time ratio for source to relay transmission in dynamic DF relay provides a trade off between throughput performance and the computational complexity of the resource allocation. The circuit power consumption has a significant impact on the the throughput of the network. 

\appendices
\section{Proof of Proposition~\ref{prop2}}
The Lagrangian dual function for the optimization problem (\ref{PS_opti_prob}), can be written as,
\begin{multline*}
D= \tau - a_1 C_1- a_2 C_2+a_3 \lambda+a_4 P_t +a_5 \tau - a_6 \big[P_d \theta \\ \quad +P_e\left(1-\theta\right)+\left(\epsilon_d+\epsilon_e\right)\tau+P_t\left(1-\theta\right) -  \eta Q g_1 \left(1-\lambda\right) \theta\big] \,
\end{multline*}
where $C_1$ and $C_2$ are the left hand sides of the constraints \eqref{constraint1} and \eqref{constraint2}, respectively. Dual variables $a_1,a_2,a_3,a_4,a_5,a_6\geqslant0$. The KKT conditions in addition to constraints (\ref{constraint1})-(\ref{constraint5}), can be written as,
\begin{equation}\label{daba_D}
\frac{\partial D}{\partial \tau}=0, \  \frac{\partial D}{\partial \lambda}=0, \ \frac{\partial D}{\partial P_t}=0 \, ,
\end{equation} 
\begin{equation}\label{comp_slack_a1}
a_1 C_1 = 0 \, , \, a_2 C_2 = 0 \, ,
\end{equation}
\begin{multline}\label{comp_slack_a6}
a_6 \big[P_d \theta  +P_e\left(1-\theta\right)+\left(\epsilon_d+\epsilon_e\right)\tau+\\P_t\left(1-\theta\right) -  \eta Q g_1 \left(1-\lambda\right) \theta\big] = 0 \, ,
\end{multline}
\begin{equation}\label{comp_slack_a3a4a5}
a_3 \lambda=0, \ a_4 P_t=0, \ a_5 \tau=0 \, .
\end{equation} 
From Lemma~1 we have that for the fixed $\bar{\theta}$, the problem is convex and satisfies Slaters condition. Thus, the KKT conditions are necessary and sufficient for the optimality. We first consider the case, when $0<\bar{\lambda}$ and $0<\bar{P_t}$. According to (\ref{comp_slack_a3a4a5}), we have $a_3=0$ and $a_4=0$. Using (\ref{daba_D}), the only possibility for $a_1$ and $a_2$ with $a_5\geqslant 0$, is $a_1>0$ and $a_2>0$. Thus, we have $a_6>0$ and from the conditions $a_1 C_1 = 0$ and $a_2 C_2 = 0$ we have $C_1=0$ and $C_2=0$, which are
\begin{equation*}\label{prop2_eq1}
\bar{\tau}=\frac{\bar{\theta}}{T_0} \ \text{log}_2 \left(1+\frac{Q g_1 \bar{\lambda}}{\left(1+\bar{\lambda}\right) \sigma^2}\right)= \frac{\left(1-\bar{\theta}\right)}{T_0}\text{log}_2 \left(1+\frac{\bar{P_t}  g_2}{2 \sigma^2}\right)
\end{equation*}
This gives $f_1\left(\bar{P_t}\right)$ and $f_2\left(\bar{P_t}\right)$ in the proposition for $\lambda$ and $\tau$, respectively. With the aid of \eqref{comp_slack_a6}, we have $f_{\bar{\theta}}\left(\bar{P_t}\right)=0$ in the proposition. By considering $f_1\left(P_t\right)$ and $f_2\left(P_t\right)$ independently, it is easily shown that $f_{\bar{\theta}}\left(P_t\right)$ is stickily decreasing with respect to $\bar{P_t}$. Moreover, $f_{\bar{\theta}}\left(c\right)\leqslant 0$ and $f_{\bar{\theta}}\left(0\right)\geqslant 0$ support that the solution to $f_{\bar{\theta}}\left(\bar{P_t}\right)=0$ is unique in $[0,c]$. 

\subsection{Lemma~1 : For fixed $\bar{\theta}$, the problem is convex. Furthermore, when $P_d \bar{\theta} +P_e\left(1-\bar{\theta}\right) <  \eta Q g_1 \bar{\theta}$ and $g_2>0$, Slater's condition is satisfied. }

Proof: By noting that $\frac{\partial^2}{\partial \lambda ^2}\text{log}_2 \left(1+\frac{Q g_1 \lambda}{\left(1+\lambda\right) \sigma^2}\right) < 0 \ ; \ \forall \lambda\geqslant 0$,
it can be used to show that the Hessian matrix is positive definite. To prove that Slater's condition holds, it is sufficient to prove that there exists a feasible $\left[\tau, \lambda, P_t\right]$ such that the equality constraint (\ref{constraint4}) is satisfied while all inequality constraints in (\ref{PS_opti_prob}) are satisfied with strict inequalities. To this end, we first parametrize $\tau, P_t, \lambda$ by a parameter $\delta \in \left[0,\eta Q g_1 \bar{\theta} - P_d \bar{\theta} - P_e\left(1-\bar{\theta}\right)\right]$ such that they satisfy the constraint (\ref{constraint4}). Thus, we can write
\begin{align*}
\tau \left(\delta\right) &= \frac{\delta}{\epsilon_d+\epsilon_e} \\
P_t\left(\delta\right) &=\frac{1}{2\left(1-\bar{\theta}\right)}\left[\eta Q g_1 \bar{\theta} -P_d \bar{\theta} - P_e\left(1-\bar{\theta}\right)-\delta\right] \\
\lambda\left(\delta\right) &=\frac{1}{2 \eta Q g_1 \bar{\theta}}\left[\eta Q g_1 \bar{\theta} -P_d \bar{\theta} - P_e\left(1-\bar{\theta}\right)-\delta\right]
\end{align*} 
The set $(0,\eta Q g_1 \bar{\theta} - P_d \bar{\theta} - P_e\left(1-\bar{\theta}\right))$ is non-empty and for any $\delta \in (0,\eta Q g_1 \bar{\theta} - P_d \bar{\theta} - P_e\left(1-\bar{\theta}\right))$ we have $P_t >0$ and $0<\lambda<1$ and $\tau>0$. Moreover, if $\delta = 0$ we have
\begin{equation*}
\frac{\bar{\theta}}{T_0} \ \text{log}_2 \left(1+\frac{Q g_1 \bar{\lambda}}{\left(1+\bar{\lambda}\right) \sigma^2}\right)>0 \, ,
\end{equation*}
and if $\delta = \eta Q g_1 \bar{\theta} - P_d \bar{\theta} - P_e\left(1-\bar{\theta}\right)$
\begin{equation*}
\frac{\bar{\theta}}{T_0} \ \text{log}_2 \left(1+\frac{Q g_1 \bar{\lambda}}{\left(1+\bar{\lambda}\right) \sigma^2}\right)=0 \, .
\end{equation*}
Thus, there exists $ \delta_1 \in (0,\eta Q g_1 \bar{\theta} - P_d \bar{\theta} - P_e\left(1-\bar{\theta}\right))$ such that
\begin{equation*}
\tau = \frac{\bar{\theta}}{T_0} \ \text{log}_2 \left(1+\frac{Q g_1 \bar{\lambda}}{\left(1+\bar{\lambda}\right) \sigma^2}\right) \ .
\end{equation*} 
Similarly, there exists $\delta_2 \in (0,\eta Q g_1 \bar{\theta} - P_d \bar{\theta} - P_e\left(1-\bar{\theta}\right))$ such that 
\begin{equation*}
\tau = \frac{\left(1-\bar{\theta}\right)}{T_0}\text{log}_2 \left(1+\frac{\bar{P_t}  g_2}{2 \sigma^2}\right) \ .
\end{equation*} 
Let $\delta_0  = \text{min}\left[\frac{\delta_1}{2},\frac{\delta_2}{2}\right]$. For all $\delta < \delta_0$, all inequality constraints in (\ref{PS_opti_prob}) are strict. This satisfies Slater's condition (Theorem 5.26 in \cite{Boyed}), which completes the proof of the lemma.


\end{document}